# Assessing the Impact of Virtualizing Physical Labs

Evgenia Paxinou, Vasilis Zafeiropoulos, Athanasios Sypsas, Chairi Kiourt, and Dimitris Kalles

**Virtual laboratories are the new online educational trend for communicating to students practical skills of science**. **In this paper we report on a comparison of techniques for familiarizing distance learning students with a 3D virtual biology laboratory, in order to prepare them for their microscopy experiment in their physical wet lab. Initial training for these students was provided at a distance, via Skype. Their progress was assessed through Pre and Post-tests and compared to those of students who opted to only prepare for their wet lab using the conventional face-to-face educational method, which was provided for all students. Our results provide preliminary answers to questions such as whether the incorporation of a virtual lab in the educational process will improve the quality of distance learning education and whether a virtual lab can be a valuable educational supplement to students enrolled in laboratory courses on Biology.**

*Learning Simulations, Virtual lab, Distance learning education, Assessment*

## I. INTRODUCTION

Laboratory skills have always been a key pillar of Natural Science Education and, almost by definition, are acquired through experience [1]. A traditional way for obtaining them is to practice in a physical lab by following the *trial-and-error* method. Unfortunately, nowadays, such a method has become time consuming and prohibitively expensive as the trainees have to interact with sensitive and expensive laboratorial equipment [2], [3], [4]. On the other hand, practicing new skills only once, and after a brief face-to face tutorial, as is usually the case in the conventional lab instruction method, can lead to low retention time of the acquired competencies and also to serious safety issues.

Having in mind that the most important factor for success in lab exercises is preparation, finding a technologically modern and convenient way for preparing students for their experiments in physical labs could be probably a robust solution for breaking the practical barriers of cost, safety and time in educational institutes [5]. Simulation based learning environments have great potentials for improving students' knowledge on scientific subjects [6], [7].

Several studies have shown that in physical labs unprepared students are preoccupied with technical and manipulative details and direct their cognitive resources towards irrelevant activities [8]. As a result, they do not get the best possible learning outcome from their experimental exercises [9]. On the contrary, by performing virtual experiments in privacy, without time and space restrictions or preoccupations with safety issues, students gain the required experience and basic information for a successful performance in subsequent physical laboratory experiments [10].

In distance learning education, students who enroll in Biology laboratorial courses can still experience the undisputable benefits of the physical lab, but less often. In this case, their preparation, in the sense of being familiarized with essential concepts and practical issues, prior to the physical lab, would be ideal.

Hellenic Open University (HOU), an institution that is mastering the distance learning education seeks new ways to communicate laboratorial skills to its distance learning students. Based on the idea that learning is an active, interpretive, iterative process [11], an interdisciplinary scientific team in HOU has recently developed a 3D game-like virtual Biology laboratory, called OnLabs. OnLabs provides a realistic 3D laboratory environment which simulates biology experiments, like light microscopy, and allows students to learn by interacting with virtual lab instruments and equipment [12].

In this paper, we are investigating the possibility of redesigning the curriculum of a Natural Sciences module by incorporating OnLabs related activities in the educational scenarios. To that end, we evaluated and compared the learning progress of three groups of students enrolled in an undergraduate program of Natural Sciences in the HOU. The students in these three groups have chosen to either participate in a Skype session, where OnLabs is used as an educational preparation tool for their microscopy exercise, or to stick to a conventional preparation.

The rest of this paper is structured in II, III and IV sections. In section II, two educational methods and the methodology of their comparison is presented in details while section III presents the results of that comparison and finally, section IV concludes the article and outlines the key directions of our future work.

Manuscript received November 14, 2017

Evgenia Paxinou is a PhD candidate at the School of Science and Technology, Hellenic Open University, Patras, Greece (tel: 00302610430360; e-mail: paxinou.evgenia@ac.eap.gr).

Vasilis Zafeiropoulos and Athanasios Sypsas are PhD candidates at the School of Science and Technology, Hellenic Open University, Patras, Greece; emails: vasiliszaf@eap.gr, sipsas@gmail.com.

Chairi Kiourt is a PhD graduate at the School of Science and Technology, Hellenic Open University, Patras, Greece; e-mail: chairik@eap.gr.

Dimitris Kalles is an Associate Professor with the School of Science and Technology, Hellenic Open University, Patras, Greece; email: kalles@eap.gr.

## II. METHODOLOGY

The sample comprised an entire class of 43 third year undergraduate distance learning students enrolled in the first cycle of a Biology laboratory course, during the summer 2017 semester, in HOU. The 43 students were randomly divided into three groups in order for each group to conduct the microscopy experiment separately, in prescheduled dates.

| No | Steps on Teaching Procedures |
|---|---|
| 1 | The biology text book is sent to all students via mail |
| 2 | PowerPoint slides are uploaded in the university platform giving additional information on microscopy |
| 3 | The 3D virtual lab OnLabs is uploaded in the university platform |
| 4 | Three weeks prior to the physical lab, *the skype-tutor* sends to all students an e-mail invitation for participation in a skype session |
| 5 | Students who are interested in the skype session, declare participation to the skype-tutor via e-mail |
| 6 | The skype-tutor sends to the participants a formal invitation for the skype session via the communication platform *Skype for Business*. The session is scheduled to take place 1-2 days prior to the physical lab |
| 7 | A week prior to the skype session, the skype-tutor sends an e-mail to the participants explaining in details the steps to be followed for connection to *Skype for Business* |
| 8 | A week prior to the skype session, the skype-tutor sends an e-mail to the participants explaining in details the steps to be followed to download OnLabs on their PC or Laptop |
| 9 | The skype-tutor urges the participants, via e-mail, to test their networks and check their connection to *Skype for Business* platform |
| 10 | The participants attend one hour skype session. In this session the skype-tutor uses PowerPoint slides to present the principles of microscopy and the parts of a photonic microscope. Finally, she performs a complete microscopy procedure using OnLabs |
| 11 | The participants are disconnected and they perform, on their own, a complete microscopy procedure using the instruction mode of OnLabs |
| 12 | The participants fill in a questionnaire to evaluate their skype experience and also OnLabs. They send the questionnaire back to the skype-tutor |
| 13 | All students appear to the biology physical lab on the scheduled date |
| 14 | All students were given a written test, the *Pre-Test*, to determine their baseline Knowledge of microscopy |
| 15 | All students attend a half an hour face-to-face tutorial on principles of microscopy |
| 16 | The lab tutor perform once, a demonstration of a complete microscopy procedure with a photonic microscope |
| 17 | All students were given a second written test, *the Post-Test*, to reassess their knowledge of microscopy. The questions in *Post-Test* are exactly the same as the ones *in Pre-Test* |
| 18 | All students fill in a questionnaire to evaluate the face-to face tutorial |
| 19 | All students uses their own photonic microscope to practice on microscopy for half an hour |
| 20 | All students fill in a work sheet to evaluate their gained experimental skills |

Fig. 1 Steps followed in traditional microscopy teaching method (1,2,3,4,13,15,16 and19) and steps followed in our enriched ,with OnLabs, educational scenario (all steps from 1 to 20). Steps 14,17,18 and 20 are the assessment steps followed by every student

Three weeks before the students' appearance in the physical lab, an e-mail invitation for survey participation was sent to all of them, via the university platform, fully explaining the project and also its research aims. Briefly, with this invitation students were asked to choose between the traditional teaching method, which includes a face-to-face tutorial and a live demonstration of a microscopy procedure, and the innovating method that includes in addition, the use of a virtual lab, of OnLabs. From the first group 7 out of 16 students were interested in using OnLabs as an extra educational tool for their preparation in microscopy, from the second group 9 out of 13 students and finally from the third group 6 out of 14 students.

In Fig. 1 we present both the steps of the traditional microscopy teaching method and the steps of the proposed educational scenario that includes using OnLabs. As Fig.1 notifies, all students followed the conventional educational method (steps 1,2,3,4,13,15,16 and 19), and the assessment procedure, (steps 14,17,18 and 20). Only those who responded to the survey invitation followed, in addition, steps 5,6,7,8,9,10,11 and 12. The *OnLabs experience* is incorporated in the new scenario, through a Skype session (step 10). The Skype environment is presented in Fig. 2.

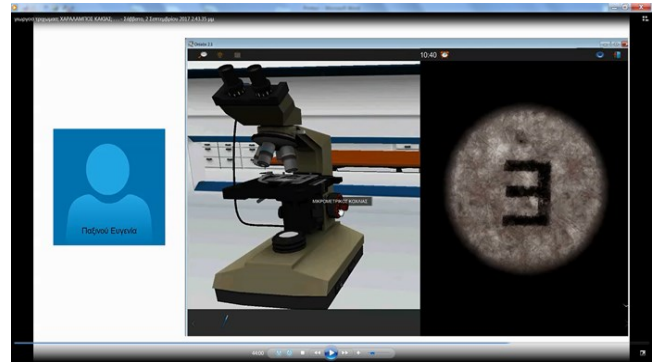

Fig. 2 Screenshot from the Skype session on microscopy, using the simulation environment of OnLabs

The objective of this study is to assess whether learning simulation could increase students' understanding of microscopy. For this assessment we took into consideration the students' grades in Pre and Post Tests (steps No 14 and 19 in Fig.1). At this point it is essential to mention that there is a fluctuation in test difficulty. Both Pre and Post-Tests given to the first group are of low difficulty, those given to the second group are of medium difficulty whereas those administered to the third group are of high difficulty. Although a microscopy expert can easily evaluate the difficulty level of a relevant test, for the estimation of the difficulty of the questions in Post and Pre Tests, we used the probabilistic approach of Item Response Theory (IRT) named Rasch Model [13]. The Rasch model uses the following probability function to estimate the probability of a student to get the question $X_j$ correct:

$$P(X_j|\theta, \beta_j) = \frac{1}{1 + e^{-(\theta - \beta_j)}}$$

where the parameter $\beta_j$ is the difficulty of a question in a test and $\theta$ is the ability of a student to answer correctly to a question of difficulty $\beta_j$. In order to use the dichotomous Rasch model we represented each question of a test as a binary variable, so that a value of 0 indicates a wrong answer and a value of 1 indicates a correct answer. For a group of $y$ students to who we administered a test of $z$ questions, we created a vector of $z$ binary variables to represent the responses given by each student. A data file in a CSV format was created containing $y$ vectors, each of size $z$. For our data analysis we used R, an open source statistical analysis language [14], and more specifically, the TAM package [15].

## III. RESULTS

The difficulty parameter $β_j$ of the IRT approach takes value in the (-∞,+∞) range, as shown in Fig. 3, with 0 indicating *Medium Difficulty*.

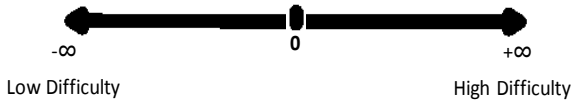

Fig. 3 The Difficulty Scale

Table I presents the means and the standard deviations of the difficulty of the questions in each test. With the Rasch model the difficulty of each question was estimated based on the students' answers. Table I provides evidence that the tests are designed in augmented difficulty from 1st to the 3rd group of students. Students' ability was also assessed but it is not shown in this study [16].

Table I: The average of the difficulty of the questions in Pre and Post-Tests in all three groups of students

|  | AVERAGE OF DIFFICULTY OF QUESTIONS | | |
| --- | --- | --- | --- |
|  | 1st Group | 2nd Group | 3rd Group |
| Pre-Test | -1.807 ± 1,277 | -2.026 ± 1,256 | 0.035 ± 1,041 |
| Post-Test | -3.064 ± 1,233 | -2.608 ± 1,440 | -0.822 ± 1,309 |

According to Table I all Post-Tests seem to be less difficult compared to their corresponding Pre-Tests, something that was expected, as Post-Tests were administered right away after the face-to-face tutorial. We used the Classical Test Theory (CTT), also known as the true score theory [17], to assess students' learning outcomes. The average of the scores in each test, and in each group, is demonstrated in Fig. 4.

Depending on Fig. 4, our first general observation is that the *With-OnLabs* students had better scores on their Pre-Tests, in all three groups, regardless of the test difficulty. This remark indicates that the *With-OnLabs* students were better prepared on microscopy. After the face-to-face tutorial the scores in Post-Tests are almost equal for all students in the 1st and 2nd group. This highlights that the knowledge given by the lab-tutor, filled successfully the knowledge gaps for the *Without-OnLabs* students when their tests were of low or medium difficulty. In the 3rd group, where both Pre and Post-Tests are the most difficult, the *With-OnLabs* students were not only better prepared with the Skype session, but had also higher scores in their Post-Tests.

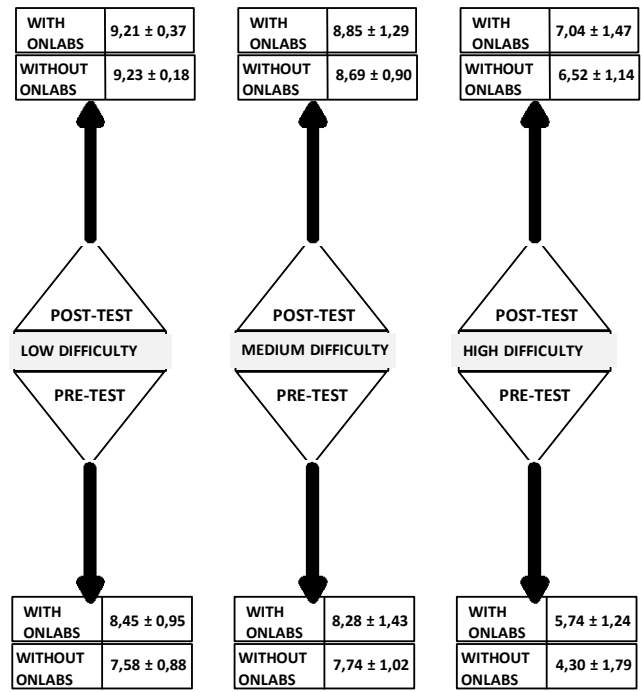

Fig. 4 Averages and standard deviations of students' score out of ten, in Pre and Post Tests of low medium and high difficulty, with or without the Skype session with OnLabs

In step No 12, (see Fig.1), the *With-OnLabs* students expressed their opinion on satisfaction, interest, confidence, understanding and cognitive load items using a 5-point Likert scale (1=Strongly Disagree, 5=Strongly Agree) in a questionnaire with names, based mostly on the ARCS-model [18]. The responses were analyzed with the statistical analysis language R and the overall conclusion drawn is that the digital learning material provides an efficient, interesting and innovative learning situation, something that is visualized in Fig.5. As it is presented in Fig.5(a) all the *With-OnLabs* students, except one, believe that a Skype session could be part of their learning procedure. Furthermore, Fig.5(b) presents that the majority of the *With-OnLabs* denotes that a Skype session should precede every exercise in the physical lab.

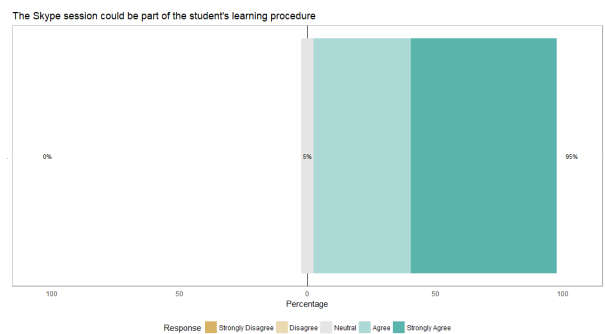

(a)

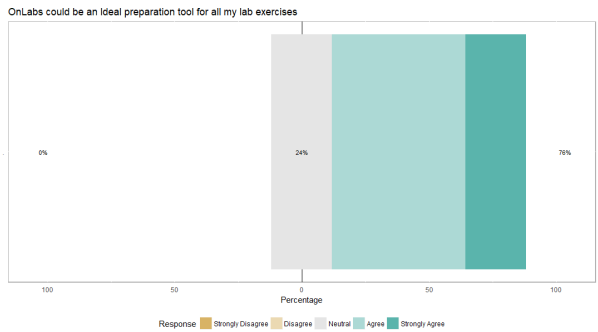

(b)

Fig. 5 *With-OnLabs* students' opinion in the following statements: (a) The Skype session could be part of the student's learning procedure; (b) OnLabs could be an ideal preparation tool for all my lab exercises.

## IV. CONCLUSIONS

In the current study, two educational methods, applied on distance learning students for their preparation on microscopy laboratorial experiment, are evaluated and compared; the conventional face-to-face lab tutorial method and our proposed educational scenario enriched with a Skype session and a 3D game-like virtual biology laboratory, called OnLabs. Our evaluation is based on the assessment of students' learning outcomes on Pre and Post-Tests. The Pre-Tests scores proved that *OnLabs experience* gave higher baseline knowledge to those students who were involved, in all groups. The Post-Tests scores showed that the face-to-face tutorial improved and equated students' understanding of concepts concerning microscopy in the 1st and 2nd group whereas in the 3rd group, where the difficulty of the tests was the highest, the *With-OnLabs* students have better grades than the *Without-OnLabs* students.

Future research should replicate the results of this study with a larger student population, in a wider selection of educational institutes and in different modules, (such further studies have been already scheduled or are in a negotiation phase). We also plan to assess the retention time of the gained knowledge in the two different educational methods, to address aspects of longer-term educational design.